\documentclass[aps,prl,twocolumn]{revtex4-1}
\usepackage{amsmath}
\usepackage{amsfonts}
\usepackage{amssymb}
\usepackage{graphicx}
\usepackage[colorlinks,linkcolor=blue,anchorcolor=blue,citecolor=blue,urlcolor=blue]{hyperref}
\usepackage{mathrsfs}
\usepackage{epsfig}
\usepackage{subfigure}
\begin{document}
\title{Microcavity-engineered plasmonic resonances for radiation enhancement and strong coupling of a quantum emitter}
\author{Pai Peng$^{1}$}
\author{Yong-Chun Liu$^{2}$}
\author{Da Xu$^{1}$}
\author{Qi-Tao Cao$^{1}$}
\author{Qihuang Gong$^{1,3}$}
\author{Yun-Feng Xiao$^{1,3}$}
\email{yfxiao@pku.edu.cn}
\altaffiliation{URL: \url{www.phy.pku.edu.cn/~yfxiao/}}
\affiliation{$^{1}$State Key Laboratory for Mesoscopic Physics and School of Physics, Peking University; Collaborative Innovation Center of Quantum Matter, Beijing 100871, People's Republic of China}
\affiliation{$^{2}$Department of Physics, Tsinghua Univeristy, Beijing 100084}
\affiliation{$^{3}$Collaborative Innovation Center of Extreme Optics, Taiyuan, 030006 Shanxi, People's Republic of China}
\date{\today}
\begin{abstract}
Localized-surface plasmon resonance is of importance in both fundamental and applied physics for the subwavelength confinement of optical field, but realization of quantum coherent processes is confronted with challenges due to strong dissipation. Here we propose to engineer the  electromagnetic environment of metallic nanoparticles (MNPs) using optical microcavities. 
An analytical quantum model is built to describe the MNP-microcavity interaction, revealing the significantly enhanced dipolar radiation and consequentially reduced Ohmic dissipation of the plasmonic modes.
As a result, when interacting with a quantum emitter, the microcavity-engineered MNP enhances the quantum yield over 40 folds and the radiative power over one order of magnitude. Moreover, the system can enter the strong coupling regime of cavity quantum electrodynamics, providing a promising platform for the study of plasmonic quantum electrodynamics, quantum information processing, precise sensing and spectroscopy.
\end{abstract}
\maketitle
Metallic nanostructures confine light on the sub-wavelength scale due to the collective excitation of electrons known as localized-surface plasmon resonances (LSPRs).
Ultrasmall mode volumes of the plasmonic resonances down to cubic nanometers make them a powerful platform for fundamental studies and novel applications, such as spaser \cite{spaser1, spaser2}, superlens \cite{superlens} and quantum plasmonics \cite{quantumplasmonics1,quantumplasmonics2}. An emerging field is to study the nanoscale light-matter interaction between plasmonic mode and few or even single quantum emitters (e.g., atoms, molecules or quantum dots, etc). For example, in the weak coupling regime LSPRs are widely used to enhance fluorescence \cite{antenna1,antenna2,fluorescence} and Raman scattering \cite{SERS3,SERS5,biosensing} and to achieve unidirectional emission \cite{unidirection}; in the strong coupling regime the coherent hybridizations between plasmonic resonances and quantum emitters have also been investigated both theoretically \cite{scth1, scth3, scth4, scth5, scth6, scth7} and experimentally \cite{scexp1, scexp2, scexp3, scexp4, scexp5, scexp6, scexp7}. 

Two approaches have been taken to achieve single-emitter strong coupling - lowering the mode volumes and suppressing the dissipation. The former typically requires ultra-fine geometries with nanometer or even subnanometer precision \cite{scexp2, scexp7}, posing challenges on fabricating metallic structures and positioning individual quantum emitters. For the latter, the dipolar plasmonic modes dissipates through both radiation and Ohmic absorption, while the multipole modes are purely absorptive \cite{Ruppin1982}.
A better coherence can be achieved by reducing the excitation of multipole modes or enhancing the excitation of dipolar modes.
Thus, efforts have been made to tailor the geometry of the metallic nanostructures, for example, elongating a metallic nanoparticle (MNP) to separate the dipolar and multipole resonances \cite{tailormnp1}, and forming dimer or arrays to obtain stronger coupling of a certain dipolar mode \cite{tailormnp2}. An alternative approach is cancelling the coupling between the emitters and the multipole modes with emitters homogeneously distributed around the metallic structure \cite{atomcloud1}.
Despite of these efforts, the nonradiative decay from the dipolar plasmonic modes remains serious for small metallic nanostructures which are advantageous for stronger light-matter interaction due to more confined fields \cite{biosensing}.
Here we provide the perspective of microcavity-engineered metallic nanostructure system, which is not revealed in previous studies of the hybrid photonic-plasmonic modes
\cite{excite5,YFX12PRA,hyPRL12,hyPRL13,hyAPL11,hyArnoldAPL11,hyBowenAPL11,hyPRB12,hyPRA13,hyPRL16,xiaoPRL10,NL10,NC12,NL15,Netherlands2,wangNL15,LPR15}.
The microcavity engineers electromagnetic environment of dipolar plasmonic mode, enhancing its radiation rate and further reducing the Ohmic absorption.
When a single quantum emitter is interacting with the cavity-engineered MNP, it shows more than 40-fold boost of quantum yield together with significant enhancement of radiative power.
With the greatly reduced dissipation, vacuum Rabi oscillation, vacuum Rabi splitting and anti-crossing phenomena arise, manifesting the system in the strong coupling regime.
The engineered dipolar plasmonic modes, together with ultrasmall mode volumes \cite{scexp2, scexp3} and suppressed excitations of multipole modes \cite{tailormnp1, tailormnp2, atomcloud1} in previous approaches, enables higher-fidelity quantum manipulation and more precise sensing. 

\begin{figure}[h]\centering
{\includegraphics[width=0.98\columnwidth]{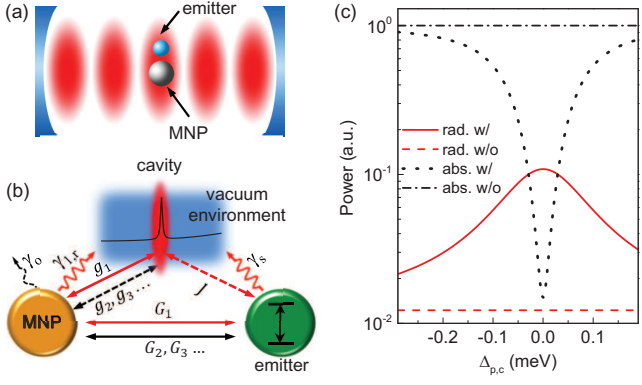}} 
\caption{(a) A quantum emitter coupled to a MNP in microcavity-engineered electromagnetic environment.
(b) Schematic of the interaction in the system. The cavity results in a peak in vacuum density of state, marked by the red region. Straight lines with arrows represent couplings and wavy lines indicate dissipations. Definitions of the symbols are clarified in the main text. The cavity-emitter coupling is tunable, and the strength $J$ vanishes by setting the direction of the emitter's dipole moment.
(c) Power outputs ($\Phi$) of MNP to each channel vs. pump-cavity detuning ($\Delta_\mathrm{p,c}$): radiation with cavity (red solid curve), radiation without cavity (red dashed curve), absorption with cavity (black dots) and absorption without cavity (black dash-dotted curve).
}
\label{fig1}
\end{figure}

We first develop an analytical model to describe the microcavity-engineered MNP system shown in Fig. \ref{fig1}(a).
The system Hamiltonian contains three parts: $H=H_\mathrm c+H_\mathrm{m}+H_{\mathrm I}$.
Here $H_\mathrm c=\omega_\mathrm c\hat c^\dag \hat c$ characterizes the cavity mode $\hat c$ with resonance frequency $\omega_\mathrm c$. $H_\mathrm m=\sum_{q,l}\omega _{q,l} \hat{a}^\dag_{q,l} \hat{a}_{q,l}$ describes the quantized plasmonic modes $\hat{a}_{q,l}$ with resonance frequencies $\omega_{q,l}$, where $q=1,2,3$ characterize three principal axes of MNP geometry and $l=1,2,3...$ labels the modes of different order (dipolar, quadrupole, etc) along the same axis. The interaction Hamiltonian reads $H_{\mathrm I}=\sum_{q,l}g_{q,l}(\hat a^\dag_{q,l} +\hat a_{q,l})(\hat c^\dag +\hat c)$, with $g_{q,l}$ being the coupling coefficient which is set to be real, shown in Fig. \ref{fig1}(b).
The quantum Langevin equations are given by
\begin{subequations}
\begin{align}
\frac{\mathrm d \hat c}{\mathrm d t}&=-\left( i\omega_c +\frac{\gamma_c}{2}\right) \hat c- i \sum_{q,l}g_{q,l} \hat a_l -\sqrt{\gamma^{\mathrm{in}}_\mathrm c} \hat c^{\mathrm{in}},\\
\frac{\mathrm d \hat a_{q,l}}{\mathrm d t}&= -\left( i\omega_{q,l} +\frac{\gamma_{q,l}}{2}\right) \hat a_{q,l}-i g_{q,l} \hat c-\sqrt{\gamma^{\mathrm{in}}_{q,l}}\hat a^{\mathrm{in}}_{q,l},
\end{align}
\end{subequations}
where $\gamma_c$ and $\gamma_{q,l}$ are the decay rates of cavity and plasmonic modes;
the inputs are introduced by $\hat c^{\mathrm{in}}$ for cavity and $\hat a^{\mathrm{in}}_{q,l}$ for plasmonic modes with rates $\gamma^{\mathrm{in}}_\mathrm c$ and $\gamma^{\mathrm{in}}_{q,l}$.

To simplify the discussion without loss of generality, hereafter we focus on spherical particles with radii much smaller than the wavelength.
Due to the spherical symmetry, only the plasmonic modes along the excitation electric field can be excited, and the footnote $q$ can be dropped.
In this case, the resonance frequencies $\omega_l$ of LSPRs can be obtained analytically using the Drude model with metallic permittivity $\epsilon_\mathrm{m }(\omega)=\epsilon _{\mathrm{\infty }}-\omega_\mathrm{p}^2/(\omega^2+i\omega\gamma_\mathrm{o})$ \cite{atomcloud1}, where $\epsilon _{\mathrm{\infty }}$ is the permittivity at high frequency, $\omega_\mathrm{p}$ is the bulk plasma frequency and $\gamma_\mathrm{o}$ is the Ohmic damping rate \cite{Drudemodel}.
For the linewidth, the dipolar mode ($l=1$) has two dissipation channels shown in Fig. \ref{fig1}(b): radiation of rate $\gamma_{1,\mathrm r}$ and Ohmic loss of rate $\gamma_\mathrm o$, with the total decay rate $\gamma_1=\gamma_{1,\mathrm r}+\gamma_\mathrm o$, while multipole modes ($l>1$) dissipate entirely through Ohmic absorption, with $\gamma_l=\gamma_\mathrm o$.
We consider a gold nanoparticle with $\epsilon _{\mathrm{\infty }}=1$, $\omega_{\mathrm p}=4~\mathrm{eV}$, $\gamma_\mathrm{o}=0.2~\mathrm{eV}$ extracted from \cite{Drude}, and the permittivity of environment is set as $\epsilon _{\mathrm{b}}=1$.
For the cavity, we choose the mode volume to be $V_\mathrm c =1~\mathrm{\mu m}^3$ and quality factor $10^5$.
We calculate the extinction spectra for the MNP-microcavity coupling system, and compare the analytical results with the numerical results computed from Mie theory. We find that (i) our analytical results agree quite well with the numerical results for MNP radius $R<30$ nm, and (ii) the coupling of the cavity mode to the multipole plasmonic modes is over 2-order of magnitude weaker than to dipolar mode \cite{SI}. Therefore, throughout this paper we omit the interaction between the multipole plasmonic modes and the cavity mode, and use a MNP with radius $R=10$ nm. Then the radiative decay rate is $\gamma_\mathrm{1,r}=2.45$ meV, calculated following Ref. \cite{Edo}. The coupling coefficient between the dipolar plasmonic mode and the cavity mode is calculated as $g_1=-2.9$ meV \cite{SI}.

To characterize the dissipative property of microcavity-engineered plasmonic excitation, we calculate the dissipation power by defining output operators. $\Phi_\mathrm r^{(i)}=\langle \hat a^{(i)\dag}_\mathrm{out,r} \hat a_\mathrm{out,r}^{(i)}\rangle$ is the output power through (i) radiation from MNP to the electromagnetic vacuum, with $\hat a^{(1)}_\mathrm{out,r}=\sqrt{\gamma_\mathrm{1,r}}\hat a_{1}$,
(ii) coupling into the cavity for further guiding with $\hat a^{(2)}_\mathrm{out,r}=\sqrt{\gamma_\mathrm c}\hat c$. $\Phi_\mathrm d=\langle \hat a^{\dag}_\mathrm{out,o} \hat a_\mathrm{out,o}\rangle$ denotes the Ohmic absorption power of dipolar plasmonic mode $\hat a_\mathrm{out,o}=\sqrt{\gamma_\mathrm o}\hat a_1$.
When the cavity and dipolar plasmonic mode are on resonance $\Delta_{\mathrm {1,c}}=\omega_\mathrm 1-\omega_\mathrm c=0$ and MNP is pumped via freespace, the total radiation and absorption output power of cavity-engineered plasmonic mode are compared to the bare plasmon case in Fig. \ref{fig1}(c). It is shown that Ohmic loss dominates over radiation in the absence of the microcavity. While in the presence of cavity, the radiation is enhanced by over one order of magnitude, and Ohmic absorption is reduced by almost 2 orders of magnitude on cavity resonance. As a result, most of the energy is guided out of the system through radiation. This phenomenon is attributed to the nontrivial electromagnetic environment created by the cavity, giving rise to a enhanced radiation of dipolar LSPR, similar to Purcell effect \cite{Purcell}.
The radiation then guides the energy out from the absorptive region, resulting in the reduction of incoherent absorption.

\begin{figure}\centering
{\includegraphics[width=0.98\columnwidth]{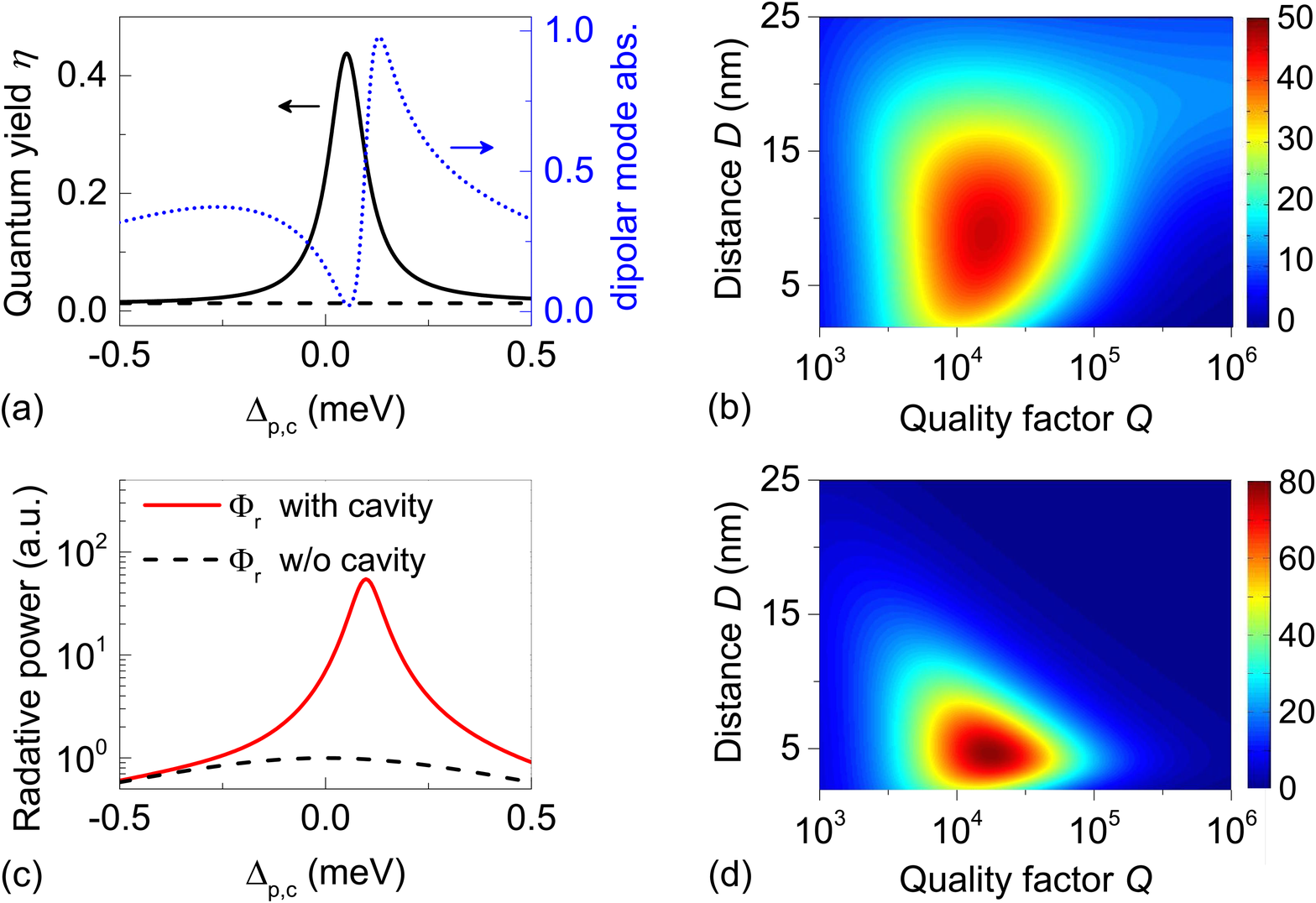}}
\caption{(a)  Quantum yield (left axis) and normalized dipolar mode absorption (right axis) vs. pump-cavity detuning $\Delta_\mathrm{p,c}$.
The black solid (dashed) curve is quantum yield with (without) the cavity. 
(b) Enhancement factor of quantum yield vs. distance $D$ and cavity $Q$ factor.
(c) Radiative power spectrum with the cavity (red solid curve) and without the cavity (black dashed curve). 
(d) Enhancement factor of total radiative power vs. distance $D$ and cavity $Q$ factor.
In (a) to (d), the parameters are \{$J$, $G$, $\gamma_\mathrm s$, $\gamma_\mathrm m$\}=\{-144, -7200, 3, 83\} $\mu$eV; the cavity, emitter and dipolar LSPR are on resonance, i.e. $\Delta_\mathrm{c,e}=\Delta_\mathrm{1,e}=0$. In (b) and (d), $\Delta_\mathrm{p,c}=\Delta_0$.
}
\label{fig2}
\end{figure}

With the property of microcavity-enhanced plasmonic radiation, we show that the system exhibits the unique advantage in coherent manipulation of a quantum emitter. After introducing an emitter, the Hamiltonian under the quantum jump approach \cite{quantumjump}
is given by
\begin{equation}\label{Ham}
H=
\begin{pmatrix}
\hat a_1^\dag\\
\hat c^\dag\\
\hat \sigma_+
\end{pmatrix}^\mathrm T
\begin{pmatrix}
\Delta_{1,\mathrm e} -i\frac{\gamma_\mathrm 1}{2}& g_1 & G\\
g_1 & \Delta_\mathrm {c,e}-i\frac{\gamma_\mathrm c}{2}  & J\\
G & J & -i\frac{\gamma_\mathrm e}{2}
\end{pmatrix}
\begin{pmatrix}
\hat a_1\\
\hat c\\
\hat \sigma_-
\end{pmatrix}.
\end{equation}
Here $\hat\sigma_-=\hat\sigma_+^\dag$ denotes the descending operator of the emitter, $G$ the coupling coefficient between the dipolar plasmonic mode and the emitter, and $J$ the coupling coefficient between the cavity and the emitter, $\Delta_{1,\mathrm e}=\omega_1-\omega_\mathrm e$ the MNP-emitter detuning and $\Delta_\mathrm {c,e}=\omega_\mathrm c-\omega_\mathrm e$ the cavity-emitter detuning. Note that the multipole plasmonic modes are far detuned from the emitter's frequency and their couplings to the emitter [$G_2, G_3 ...$ in Fig. \ref{fig1}(b)] are weaker than the dipolar plasmonic mode, so it is reasonable to treat them as a heat bath interacting with the emitter, well characterized by the Markovian approximation \cite{SI}. As a result, the decay rate of the emitter is $\gamma_\mathrm e=\gamma_\mathrm s+\gamma_\mathrm m$, consisting of both radiation to vacuum with rate $\gamma_\mathrm s$ and dissipation to multipole plasmonic modes with rate $\gamma_\mathrm m$ \cite{SI}.

With a weak pump to the emitter, we study the quenching effect by calculating the quantum yield
\begin{equation}
\eta=\frac{\sum_i\Phi_\mathrm r^{(i)}}{\sum_i \Phi_\mathrm r^{(i)}+\sum_j\Phi_\mathrm o^{(j)}}.
\end{equation}
In comparison with the previous case, the presence of a quantum emitter changes the output channels. Both the emitter and MNP radiate to vacuum, so $\hat a^{(1)}_\mathrm{out,r}=\sqrt{\gamma_\mathrm{1,r}}\hat a_{1}+\sqrt{\gamma_\mathrm s}\hat \sigma_-$. Absorption through multipole modes is accounted for by an additional Ohmic output operator $\hat a^{(2)}_\mathrm{out,o}=\sqrt{\gamma_\mathrm m}\sigma_-$.
In Fig. \ref{fig2}(a) we plot the quantum yield for an emitter with dipole moment $\mu=1$ e$\cdot$nm and the distance between the emitter and the MNP surface being $D=10$ nm. It shows that the quantum yield reaches over $40\%$ in the presence of a cavity, while it is only $1\%$ for a bare MNP case.
This significant enhancement of quantum yield results from microcavity-reduced plasmonic absorption, demonstrated by the blue dotted curve in Fig. \ref{fig2}(a).
The maximum quantum yield corresponds to the valley of the Fano-lineshape absorption spectrum of the dipolar mode. Note that the Fano resonance originates from interference of two channels that excite the dipolar modes - (i) direct excitation from the quantum emitter and (ii) indirect excitation from the emitter through cavity to dipolar mode, see Fig. \ref{fig1}(b). Maximum destructive interference is achieved for
\begin{equation}
\Delta_\mathrm{p,c}=-J\frac{g_1}{G}\equiv\Delta_0.
\end{equation}
The peak of MNP excitation does not correspond to the minimum of quantum yield because both the Ohmic loss and the radiation power are large. Although the excitation of MNP is reduced at $\Delta_\mathrm{p,c}=\Delta_0$, compromising the local density of state near the quantum emitter, radiation enhancement due to the cavity environment still leads to over one-order-of-magnitude larger output power, as shown in Fig. \ref{fig2}(c).

Figures \ref{fig2}(b) and (d) plot, respectively,  the enhancement factors of quantum yield and radiative power vs. emitter-MNP distance $D$ and cavity $Q$ factor. It is counterintuitive that engineering effect of the microcavity does not increase monotonically with increasing $Q$, instead it has a maximum at $Q^\mathrm{opt}$. This phenomena is attributed to two counteracting effects: (i) when $Q<Q^\mathrm{opt}$, absorption of dipolar plasmonic mode dominates the energy dissipation, because the electromagnetic environment of a broad cavity resonance is basically the same as vacuum; (ii) when $Q>Q^\mathrm{opt}$, although the dipolar plasmonic mode dissipates most of its energy through radiation, the energy guiding channel through the cavity becomes less efficient for a high $Q$ factor, so that a substantial portion of energy is absorbed by multipole modes. Balance of the two effects leads to a maximum radiation enhancement at $Q=Q^\mathrm{opt}$. Figures \ref{fig2}(b) and (d) also show that the effect of cavity engineering also depends on the emitter-MNP distance $D$.
When $D$ is small, the loss from the multipole plasmonic modes dominates; for a large $D$, the coupling between the emitter and LSPRs becomes weaker. In the present case, the over 20-fold radiation enhancement can be achieved in a wide range $5~\mathrm{nm}<D<15~\mathrm{nm}$.

\begin{figure}[t]\centering
{\includegraphics[width=1\columnwidth]{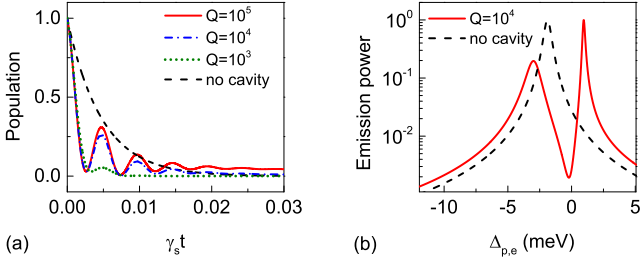}}
\caption{
(a) Temporal evolution of the population on the emitter for $Q=10^5$ (red solid curve), $Q=10^4$ (blue dash-dotted curve), $Q=10^3$ (green dots) and without cavity (black dashed curve).
(b) Radiation power of the emitter with respect to pump-emitter detuning $\Delta_\mathrm{p,e}$ for $Q=10^4$ (red solid curve) and without cavity (black dashed curve).
In (a) and (b), $\theta=60^\circ$, emitter-MNP vertex distance $D=5$ nm, cavity mode volume $0.1~\mathrm{\mu m}^3$, detunings $\Delta_{1,\mathrm e}=0.6$ eV and $\Delta_\mathrm{c,e}=1.5$ meV.
}
\label{fig3}
\end{figure}

\begin{figure}[t]
{\includegraphics[width=0.98\columnwidth]{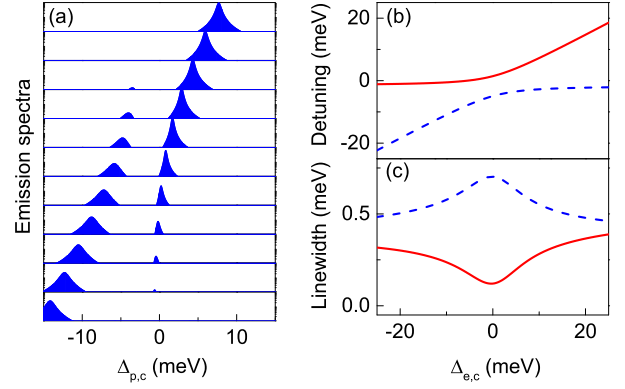}}
\caption{
(a) Emission spectra (in logrithmic scale) of the emitter with respect to the pump-cavity detuning $\Delta_\mathrm{p,c}$ for various emitter-cavity detunings $\Delta_\mathrm{e,c}$. From bottom to top,  $\Delta_\mathrm{e,c}$ increases from -10 meV to 10 meV with a 2 meV step.
Detunings (b) and linewidths (c) of the strongly coupled cavity-engineered plasmonic mode and the emitter. 
}
\label{fig4}
\end{figure}

The reduction of incoherent absorption is able to bring the originally weak-coupled emitter-MNP system into the strong coupling regime. Different from the previous semi-classical treatment of the hybrid photonic-plasmonic mode \cite{YFX12PRA, hyPRL12, hyPRL13} 
, here we provide the full quantum model which is straight forward to study the strong coupling dynamics. To emphasize that the role of cavity is to engineer MNP, rather than to couple the emitter directly, we deliberately design the geometry of the system by setting the dipole moment of the emitter $\vec{\mu}$ perpendicular to cavity field polarization $\vec{E}_\mathrm{cav}$, so that cavity-emitter coupling $J=0$.
A ellipsoidal MNP with semi-principal axes $(33, 5.5, 5.5)$ nm is used here, whose long axis lies in the plane formed by $\vec{\mu}$ and $\vec{E}_\mathrm{cav}$.
Therefore the dipolar plasmonic mode of long axis couples to both the cavity and the emitter, with compromised coefficients $G(\theta)=G\cos\theta$ and $g_1(\theta)=g_1\sin\theta$, where $\theta$ is the included angle between the long axis and $\vec{E}_\mathrm{cav}$.
The cavity-engineered system exhibits the vacuum Rabi oscillation and splitting, the typical manifests of strong coupling, plotted in Figs. \ref{fig3}(a) and (b). In the temporal domain, the emitter undergoes 5 complete periods of oscillation for $Q=10^5$ [Fig. \ref{fig3}(a)]. As the quality factor decreases, the number of periods declines, and oscillation does not occur without the cavity, demonstrating that the cavity engineering is crucial for achieving strong coupling. In the frequency domain, the system exhibits two well-separated peaks in the emission spectra of emitter for $Q=10^4$, and the corresponding splitting reaches $4$ meV [Fig. \ref{fig3}(b)]. The emission spectra are further illustrated in Fig. \ref{fig4}(a) with different emitter-cavity detuning $\Delta_\mathrm{e,c}$. The anti-crossing shown in the spectra is studied by analyzing the eigenfrequencies and linewidths of the two involved eigenstates, plotted in Fig. \ref{fig4}(b) and (c). At $\Delta_\mathrm{e,c}=0$, the energy level splitting $2g_\mathrm{eff}$ reaches $3.5$ meV, larger than linewidths of both eigenstates $\kappa_1=1.28$ meV and $\kappa_2=0.11$ meV. The cooperativity $4g_\mathrm{eff}^2/(\kappa_1\kappa_2)$ exceeds $80$. The linewidths do not show a crossed pattern as in the conventional cavity-QED case \cite{quantumoptics},
because cavity mediated LSPR-emitter coupling poses both dispersive and dissipative nature. The similar mechanism was reported in exciton mediated optomechanical coupling \cite{mediatedcoupling1} and low-$Q$ cavity mediated cavity-QED coupling \cite{LYC14PRL}. The dispersive coupling leads to the repulsion of eigenfrequencies' real parts [Fig. \ref{fig4}(b)] and the dissipative coupling results in repulsion of the imaginary parts [Fig. \ref{fig4}(c)].
This strong light-matter interaction between a single emitter and LSPRs possesses potential for various application, such as efficient transfer information from emitter to photon and high-fidelity manipulation of qubit.

In summary, we propose to engineer the electromagnetic environment of a MNP using a microcavity. By constructing an analytic quantum model to describe the MNP-microcavity interaction, we find that the microcavity enhances the radiation of the dipolar plasmonic mode, and thus reduces the Ohmic dissipation of the MNP. When interacting with a quantum emitter, the microcavity-engineered MNP system shows more than $40$-fold enhancement of the quantum yield and greatly increases the radiative power output. Most importantly, with the dissipation suppressed, the microcavity-engineered plasmonic system can reach the strong coupling regime of cavity quantum electrodynamics, while the bare MNP-emitter interaction falls into the weak coupling regime. We envision that the cavity-engineered plasmonic system holds great potential for the study of nanoscale cavity quantum electrodynamics and precise sensing.

\begin{acknowledgments}
We thank Xi Chen and Linbo Shao at Harvard University for fruitful discussion. 
This project was supported by the Ministry of Science
and Technology of China (Grants No. 2016YFA0301302,
No. 2013CB921904, and No. 2013CB328704) and the NSFC
(Grants No. 61435001, No. 11654003, and No. 11474011).
P.P. was supported by the National Fund for Fostering Talents of Basic Science. 
\end{acknowledgments}

\emph{Note added.} - While finishing this letter, we noted an independent work on the arXiv that studied the similar system using numerical simulation and the quasinormal mode approximation. Ref. \cite{vahid}.

\end{document}